\documentclass[]{aastex62}
\usepackage{CJK}

\received{March 11, 2019}
\revised{January 29, 2020 }
\accepted{January 30, 2020 }
\published{March 6, 2019}

\submitjournal{ApJ on March 9, 2019}
\shorttitle{An empirical candidate zone for BL Lacs}
\shortauthors{Kang et al.}

\begin{document}
 \begin{CJK*}{UTF8}{gbsn}

\title{An empirical ``high-confidence" candidate zone for $Fermi$ BL Lacertae Objects}


\correspondingauthor{Shi-Ju Kang}
\email{kangshiju@alumni.hust.edu.cn}

\author[0000-0002-9071-5469]{Shi-Ju Kang}
\affil{School of Physics and Electrical Engineering, Liupanshui Normal University, Liupanshui, Guizhou, 553004, People's Republic of China}
\affiliation{Guizhou Provincial Key Laboratory of Radio Astronomy and Data Processing, Guiyang, Guizhou, 550001, People's Republic of China}

\author{Kerui Zhu}
\affiliation{Department of Physics, Yunnan Normal University, Kunming, Yunnan, 650092, People's Republic of China}

\author{Jianchao Feng}
\affiliation{School of Physics and Electronic Science, Guizhou Normal University, Guiyang, Guizhou, 550001, People's Republic of China}
\affiliation{Guizhou Provincial Key Laboratory of Radio Astronomy and Data Processing, Guiyang, Guizhou, 550001, People's Republic of China}
\collaboration{}

\author[0000-0003-4773-4987]{Qingwen Wu}
\affiliation{School of Physics, Huazhong University of Science and Technology, Wuhan, Hubei, 430074, People's Republic of China}

\author[0000-0003-4111-5958]{Bin-Bin Zhang}
\affiliation{School of Astronomy and Space Science, Nanjing University, Nanjing 210093, People's Republic of China}
\affiliation{Key Laboratory of Modern Astronomy and Astrophysics, Nanjing University, Ministry of Education, People's Republic of China}
\affiliation{Department of Physics and Astronomy, University of Nevada Las Vegas, NV 89154, USA}

\author{Yue Yin}
\affiliation{School of Physics and Electrical Engineering, Liupanshui Normal University, Liupanshui, Guizhou, 553004, People's Republic of China}

\author{Fei-Fei Wang}
\affiliation{School of Mathematics and Physics, Qingdao University of Science and Technology, Qingdao, Shandong, 266061, People's Republic of China}

\author{Yu Liu}
\affiliation{School of Physics, Huazhong University of Science and Technology, Wuhan, Hubei, 430074, People's Republic of China}

\author{Tian-Yuan Zheng}
\affiliation{The High School Affiliated to Anhui Normal University, Wuhu, Anhui, 241000, People's Republic of China}

\begin{abstract}

In the third catalog of active galactic nuclei detected by the $Fermi$ Large Area Telescope Clean (3LAC) sample, there are 402 blazars candidates of uncertain type (BCU). The proposed analysis will help to evaluate the potential optical classification flat spectrum radio quasars (FSRQs) versus BL Lacertae (BL Lacs) objects of BCUs, which can help to understand which is the most elusive class of blazar hidden in the Fermi sample. By studying the 3LAC sample, we found some critical values of  $\gamma$-ray photon spectral index ($\Gamma_{\rm ph}$), variability index (VI) and radio flux (${\rm F_R}$) of the sources separate known FSRQs and BL Lac objects. We further utilize those values to defined an empirical ``high-confidence" candidate zone that can be used to classify the  BCUs. Within such a zone ($\Gamma_{\rm ph}<2.187$, log${\rm F_R}<2.258$ and ${ \rm logVI <1.702}$), we found that 120 BCUs can be classified BL Lac candidates with a higher degree of confidence  (with a misjudged rate $<1\%$).  Our results suggest that an empirical ``high confidence" diagnosis is possible to distinguish the BL Lacs from the Fermi observations based on only on the direct observational data of $\Gamma_{\rm ph}$, VI and ${\rm F_R}$.

\end{abstract}


\section{Introduction} \label{sec:intro}

Blazars are a particular class of radio-loud active galactic nuclei (AGNs) with a relativistic jet pointing toward the Earth. The broadband (from radio up to TeV energies) emissions of blazars are mainly dominated by non-thermal components which are produced in the relativistic jet \citep{1995PASP..107..803U}. According to the strength of the optical spectral lines, blazars can be further divided into two subclasses \citep{1991ApJ...374..431S,1991ApJS...76..813S,1999ApJ...525..127L}, namely, the flat spectrum radio quasars (FSRQs; strong  emission lines with equivalent width EW $\ge{\rm 5 \AA}$ in rest frame) and the BL Lacerate objects (BL Lacs; weak or no emission and absorption lines). The multi-wavelength spectral energy distributions (SEDs) from the radio to $\gamma$-ray bands of normally exhibit a two-hump structure in the ${\rm log \nu-log \nu F_{\nu}}$ space. The low energy bump (peaking between millimeter and soft X-ray range)  is always explained as synchrotron emission from the non-thermal electrons in the relativistic jet, while the high energy bump (peaking within MeV-GeV energy range) is inverse Compton (IC) scattering. Furthermore, based on the peak frequency ($\nu^{\rm S}_{\rm p}$) of the lower energy bump, blazars can also be classified as low- ( $\nu^{\rm S}_{\rm p}<10^{14}$ Hz), intermediate-( $10^{14}~\rm Hz<\nu^{\rm S}_{\rm p}<10^{15}$ Hz) and high-( $\nu^{\rm S}_{\rm p}>10^{15}$ Hz ) synchrotron-peaked sources \citep[i.e., LSPs, ISPs, and HSPs, ][]{2010ApJ...716...30A}.

This work utilizes the third catalog of AGNs detected by the $Fermi$-LAT (3LAC) sample \citep{2015ApJ...810...14A}, which is part of the first four years of the $Fermi$-LAT data, the third $Fermi$ Large Area Telescope (LAT) source catalog \cite[3FGL,][]{2015ApJS..218...23A}.
 The 3LAC clean sample (i.e., the high-confidence clean sample of the 3LAC) reports 1444 $\gamma$-ray AGNs: 414 FSRQs ($\sim$~30\%), 604 BL Lac objects ($\sim$~40\%), 402 blazar candidates of uncertain type (BCU, $\sim$~30\%) and 24 non-blazar type AGNs ($<$~2\%) \citep{2015ApJ...810...14A}. The identification of FSRQs and BL Lacs are solid, mostly based on the clear evidence on the (non-)existing of emission and/or absorption lines.
On the other hand, BCUs are those sources without a confirmed classifications due to the missing representative features on optical spectrum (BCU I), synchrotron peak frequencies of SED (BCU II), and/or their broadband emissions (BCU III) (see \citealt{2015ApJ...810...14A,2015ApJS..218...23A} for the details and references therein).
Studying such a large sample of BCUs is crucial to understand of the physics of $\gamma$-ray emission of blazars
(e.g., \citealt{2012ApJ...753...45S,2015MNRAS.454..115S,2016ApJS..226...20F,2018RAA....18...56K,2019ApJ...872..189K,2019ApJ...887..134K,2020arXiv200106010Z}).

Estimating the possible classification BL Lac vs FSRQ of BCUs can help to understand which is the most elusive class of blazar hidden in Fermi sample (\citealt{2015ApJS..217....2M}). 
Indeed, some potential BL Lac or FSRQ candidates can be identified from the BCUs sample in the 2FGL/3FGL catalogues using different approaches such as supervised machine learning (e.g., support vector machine [SVM] and random forest [RF]; \cite{2013MNRAS.428..220H}), neural network \citep{2016MNRAS.462.3180C}, artificial neural network \citep[ANN;][]{2017MNRAS.470.1291S}, multivariate classification method \citep{2017A&A...602A..86L}, and by statisical analysis of the broadband spectral properties \citep[including spectral indices in the gamma-ray, X-ray, optical, and radio bands;][]{2017ApJ...838...34Y}.
In addition, we've identified potential BL Lacs and FSRQs candidates from the 3LAC Clean sample using 4 different SML algorithms (Mclust Gaussian finite mixture models, Decision trees, RF, and SVM; {\citealt{2019ApJ...872..189K} [Paper I]}) and from the 4FGL catalogue using 3 different SML algorithms (ANN, RF, and SVM; \citealt{2019ApJ...887..134K}).
Nevertheless, the final confirmation of the BCU nature of candidates in all above approaches is subject to the observations of optical spectroscopy or counterparts in other wavelength (e.g., \citealt{2014AJ....148...66M};
\citealt{2016Ap&SS.361..316A,2016AJ....151...32A,2016AJ....151...95A};
\citealt{2016Ap&SS.361..337M};
\citealt{2016A&A...596A..10M};
\citealt{2017MNRAS.467.2537K};
\citealt{2017Ap&SS.362..228P};
\citealt{2018AJ....156..212M};
\citealt{2019ApJS..241....5D}; 
\citealt{2019Ap&SS.364....5M};
\citealt{2019Ap&SS.364...85P}),
or broadband spectral features (e.g., Fermi/LAT collaboration, \citealt{2009A&A...495..691M,2012ApJ...750..138M,{2016Ap&SS.361..337M},2016Ap&SS.361..316A,2016AJ....151...32A,2016AJ....151...95A}).
If such information is missing, classification of BCUs will become challenging especially when no training set is available
(see e.g., \citealt{2013ApJ...764..135S}; \citealt{2015AJ....149..163L}; \citealt{2015AJ....149..160R}; \citealt{2017ApJ...851..135P,2017FrASS...4...45P}; \citealt{2018ApJ...861..130L}; \citealt{2019ApJ...871...94K}; \citealt{2019ApJ...871...94K}). To overcome such difficulties, in this letter, we aim to evaluate the potential classification of BCUs based on only on the direct observational properties in $\gamma$-ray and radio band. Such properties include $\gamma$-ray photon spectral index ($\Gamma_{\rm ph}$), and variability index (VI) and radio flux (${\rm F_R}$). By perform some detailed analysis, we confirmed the existence of a high-confident zone where the condition-met BCUs are most likely BL Lac objects.

We organize the present paper as follows.
In Section 2, a brief description on the sample selection is provided followed by the proposed analysis methods and results.
Comparisons of between our results with some other recent results are presented in Section 3.
Our results are discussed in Section 4 and summarized in Section 5.

\begin{figure*}[tbp]
\centering
\includegraphics[height=20.8cm,width=18cm]{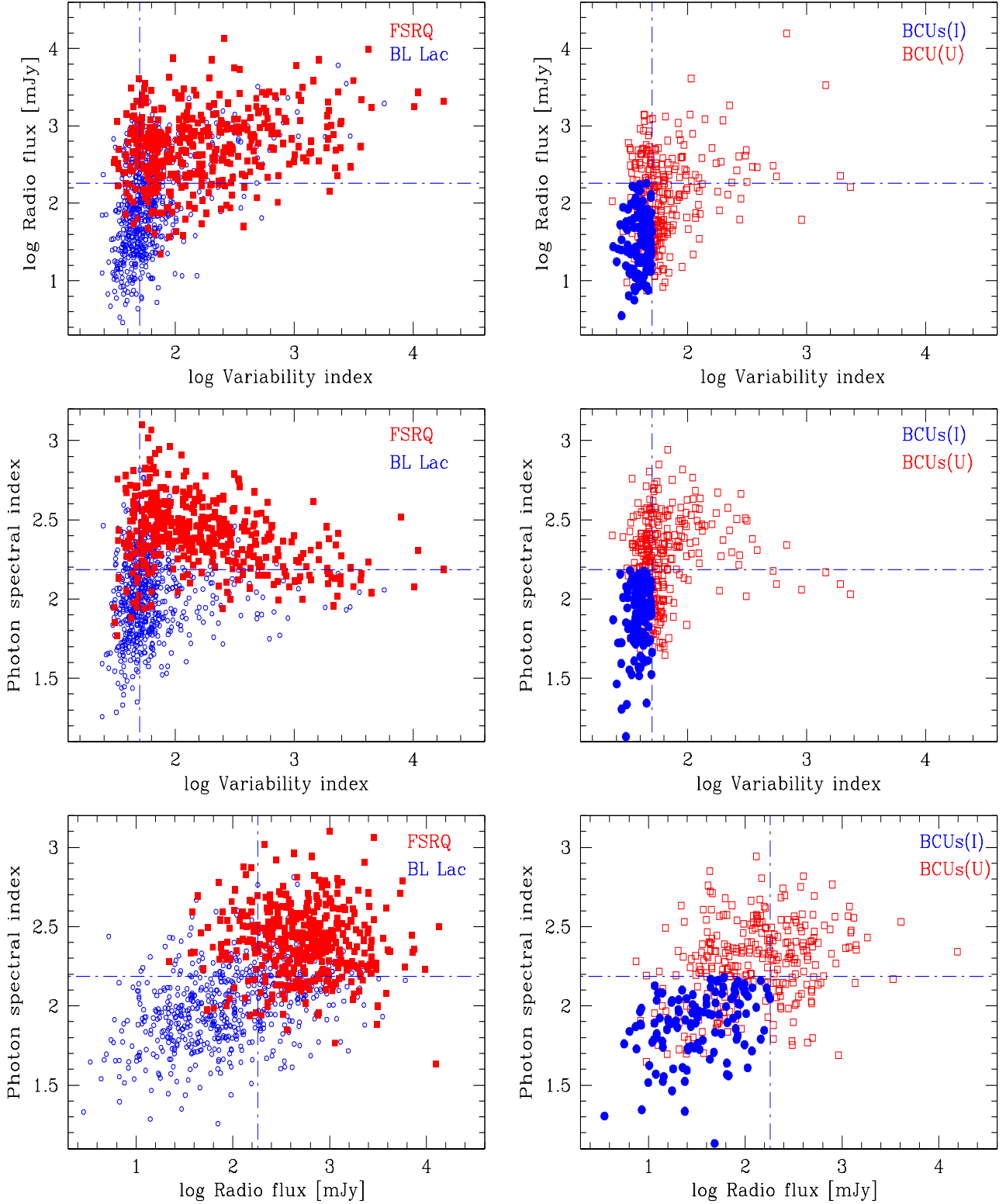}
\caption{(Color online) The scatterplots of the variability index (VI), radio flux (${\rm F_R}$) and photon spectral Index ($\Gamma_{\rm ph}$) for fermi blazars (left column), where red solid squares represent FSRQs and blue empty points represent BL Lacs. The right panels represents the scatterplots of the BCUs (right column), where the BCUs (I) --- are the identified BL Lacs (blue solid points) using the ``$a_{i} < X$ zone" and the BCUs (U) --- are the unidentified BCUs (red empty squares). The dotted-dashed parallel and perpendicular blue lines indicate $\Gamma_{\rm ph}$, log${\rm F_R}$, and log${\rm VI}$ is equal to 2.187, 2.258, and 1.702, respectively.
 }\label{sub:fig_all}
\end{figure*}

\begin{table*}[tbp]
	\centering
	\caption{The result of Two-sample Test}
	\label{tab:test}
	\begin{tabular}{clcccccccccchh} 
		\hline\hline
	&Using &\multicolumn{2}{c}{{KS test}} & &\multicolumn{3}{c}{t-test} & &\multicolumn{2}{c}{Wilcoxon-test} \\
\cline{3-4} \cline{6-8}	 \cline{10-11}			
{test dataset}&{Paramaters} & {$D$} & {$p_{1}$}&~& {$t$} &df& {$p_{2}$} &~& {$W$} & {$p_{3}$} & &{$\eta_{\rm test}^{'}$}\\
		\hline
 &$\Gamma_{\rm ph}$, ${ \rm log VI}$, log${\rm F_R}$ 	 &0.514 &0 &&32 	&2455 & {$<$1.0E-6}	  &&1826100 & {$<$1.0E-6} &&$\sim$0.132\\
604 BL Lacs&$\Gamma_{\rm ph}$, log${\rm F_R}$ 	 &0.587 &0 &&31 	&1915  & {$<$1.0E-6} 	  &&850970  & {$<$1.0E-6} &&$\sim$0.147\\
vs. &$\Gamma_{\rm ph}$, log${\rm VI}$                 	 &0.497 &0 &&25 	&1399  & {$<$1.0E-6} 	  &&803250 & {$<$1.0E-6}  &&$\sim$0.137\\
414 FSRQs &${ \rm log VI}$, log${\rm F_R}$            	 &0.490 &0 &&25 	&1596  & {$<$1.0E-6} 	  &&802600 & {$<$1.0E-6}  &&$\sim$0.235\\
 &$\Gamma_{\rm ph}$                                            	 &0.627 &0 &&27 	&955   & {$<$1.0E-6}  	  &&222270 & {$<$1.0E-6}  &&$\sim$0.137\\
 &log${\rm F_R}$                                                     	 &0.562 &0 &&23 	&996   & {$<$1.0E-6} 	  &&210750 & {$<$1.0E-6}   &&$\sim$0.265\\
 &log${\rm VI}$                                                        	 &0.478 &0 &&15 	&608   & {$<$1.0E-6} 	  &&197720 & {$<$1.0E-6}   &&$\sim$0.309\\
		\hline
 &$\Gamma_{\rm ph}$, ${ \rm log VI}$, log${\rm F_R}$ 	 &0.725 &0 &&38 	&926 & {$<$1.0E-6}          &&416470 & {$<$1.0E-6} &&0.0\\ 
 120 BL Lacs &$\Gamma_{\rm ph}$, log${\rm F_R}$ 	 &0.849 &0 &&32 	&427 & {$<$1.0E-6}  	 &&192930 & {$<$1.0E-6} &&0.0\\
 vs. &$\Gamma_{\rm ph}$, log${\rm VI}$                        &0.659 &0 &&29 	&770 & {$<$1.0E-6}          &&179540 & {$<$1.0E-6} &&0.0\\
414 FSRQs &${ \rm log VI}$, log${\rm F_R}$              	 &0.753 &0 &&35 	&803 & {$<$1.0E-6}  	 &&188340 & {$<$1.0E-6} &&0.0\\
 &$\Gamma_{\rm ph}$                                                 &0.848 &0 &&24    &198 & {$<$1.0E-6}  	 &&48137   & {$<$1.0E-6} &&0.0\\ 
 &log${\rm F_R}$                                                    	 &0.862 &0 &&28    &231 & {$<$1.0E-6}   	 &&48580   & {$<$1.0E-6} &&0.0\\ 
 &log${\rm VI}$                                                       	 &0.884 &0 &&24 	&461 & {$<$1.0E-6}   	 &&47636   & {$<$1.0E-6} &&0.0\\
 		\hline
	\end{tabular}\\
\tablecomments{Column 1 shows the test dataset: 604 BL Lacs vs. 414 FSRQs, or 120 
BL Lac candidates vs. 414 FSRQs. 
Column 2 shows the parameters satisfied simultaneously used in the test. 
Column 3 and Column 4 give the value of the test statistic ($D$) and the p-value ($p_1$) for the two-sample Kolmogorov$-$Smirnov test;
The value of the t-statistic ($t$), the degrees of freedom for the t-statistic (df) and the p-value ($p_2$) for the Welch Two Sample t-test are listed in Column 5, Column 6 and Column 7 respectively; 
Column 8 and Column 9 report the value of the test statistic ($W$) and the p-value ($p_3$) for the Wilcoxon rank sum test with a continuity correction.
All data are obtained by R code ({\url{https://www.r-project.org/}}) (see \citealt{RLanguage}).
}
\end{table*}

\begin{table*}[tbp]
	\centering
	\caption{The result of One-sample test}
	\label{tab:onetest}
	\begin{tabular}{clcccccccccch} 
			\hline\hline
	&Using &\multicolumn{2}{c}{{KS test}} & &\multicolumn{3}{c}{t-test} & &\multicolumn{2}{c}{Wilcoxon-test} \\
\cline{3-4} \cline{6-8}	 \cline{10-11}		
{test dataset}&{Paramater} & {$D$} & {$p_{1}$}&~& {$t$} &df& {$p_{2}$} &~& {$W$} & {$p_{3}$} & &{$\eta_{\rm test}^{'}$}\\
		\hline
 &$\Gamma_{\rm ph}$         &0.964 &0 &&226 	&413 &0                                   	 &&85905 &{$<$1.0E-6}&&$\sim$0.137\\
414 FSRQs &log${\rm F_R}$ &0.942 &0 &&118       &413 &{$<$1.0E-6}	 &&85905 &{$<$1.0E-6}&&$\sim$0.265\\
 &log${\rm VI}$                   &0.931 &0 &&85    	&413 &{$<$1.0E-6}	 &&85905 &{$<$1.0E-6}&&$\sim$0.309\\
		\hline
	\end{tabular}\\
\tablecomments{Column 1 shows the test dataset: the 3 parameters of the 414 FSRQs.
Column 2 shows the parameter used in the one-sample test. 
Column 3 and Column 4 give the value of the test statistic ($D$) and the p-value ($p_1$) for the One-sample Kolmogorov$-$Smirnov test;
The value of the t-statistic ($t$), the degrees of freedom for the t-statistic (df) and the p-value ($p_2$) for the Welch One Sample t-test are listed in Column 5, Column 6, and Column 7, respectively; 
Column 8 and Column 9 report the value of the test statistic ($W$) and the p-value ($p_3$) for the Wilcoxon signed rank test with a continuity correction.
All data are obtained by R code ({\url{https://www.r-project.org/}}) (see \citealt{RLanguage}.)
}
\end{table*}

\section{Data Selection and Analysis }\label{sec:sample}

We selected the data from the 3LAC Clean sample\footnote{\url{http://www.asdc.asi.it/fermi3lac/}} in the 3FGL Catalog\footnote{\url{https://fermi.gsfc.nasa.gov/ssc/data/access/lat/4yr_catalog/}}. In order to perform the analysis, we selected the sources with available measurements of $\Gamma_{\rm ph}$, VI, and ${\rm F_R}$, which yields to a sample of 1418 Fermi blazars, including 414 FSRQs, 604 BL Lacs and 400 BCUs (two sources that have no radio data are excluded)

In order to investigate whether there is a characteristic zone in the 3-parameter (namely, $\Gamma_{\rm ph}$, VI, and ${\rm F_R}$), we first exhibited the scatterplots of the known FSRQs and BL Lacs samples. In Figure \ref{sub:fig_all}, the 2-D scatterplots between any two parameters of $\Gamma_{\rm ph}$, VI, and ${\rm F_R}$ for the identified FSRQs and BL Lacs are shown in the left column. One can immediately notices that the values of $\Gamma_{\rm ph}$, ${\rm VI}$, and ${\rm F_R}$ of the FSRQs are normally larger than those of the BL Lacs. FSRQs feature comparatively concentrated distribution, while the BL Lacs show a relatively wider distribution.
The distributions of $\Gamma_{\rm ph}$, ${\rm VI}$, and ${\rm F_R}$ between the FSRQs and BL Lacs groups exhibit significantly different behavior.
The two-sample Kolmogorov$-$Smirnov test for these 3 parameters gives the value of the test statistic D = 0.514 and the p-value $ p_1 = 0$;
the Welch Two Sample t-test gives the value of the t-statistic t = 32, the degrees of freedom for the t-statistic df = 2455 and the p-value $ p_2 < $1.0E-6; 
while the Wilcoxon rank sum test with the continuity correction gives the value of the test statistic W = 1826100 and p-value $ p_3 < $1.0E-6 (obtained by R\footnote{\url{https://www.r-project.org/}} code \citealt{RLanguage})
for all the 3 parameters (see Table \ref{tab:test}). 
For the other parameter combinations, either one or two parameters, the test results are also listed in Table \ref{tab:test}. 
The results significantly reject the hypothesis that the two distributions (FRSQs, BL Lacs) are drawn from the same distribution.

We find the two samples (marked red and blue) can be well separated by some critical lines with the following value: $\Gamma_{\rm ph}$=2.187, log${\rm F_R}$=2.258, and log${\rm VI}$=1.702. The three critical values are obtained in the following procedure: 
\begin{enumerate}
\item We performed a one-sample normal distribution test (e.g., KS-test, t-test, and Wilcoxon test) for the $\Gamma_{\rm ph}$, ${\rm VI}$, and ${\rm F_R}$ of the FSRQs and found that the distribution of those parameters are consistent with a Gaussian distribution with significant p-value (Table \ref{tab:onetest}). 
\item We further obtained lowest one-sided confidence interval value ($a_{1}=2.187$, $a_{2}=2.258$, and $a_{3}=1.702$) under the assumptions that $\Gamma_{\rm ph}$, ${\rm VI}$, and ${\rm F_R}$ of FSRQs are normally distributed, which are assigned as the critical value mention above.
\end{enumerate}

We find that there are no FSRQs falling in a range $\Gamma_{\rm ph} < a_{1}$, log${\rm F_R} < a_{2}$ and ${ \rm logVI} < a_{3}$, 
while some BL Lacs lie in the zone ($a_{i} < X$), where $a_{i}$ (i=1,2,3) is set as the boundary value. 
Moreover, there are only 3 FSRQs in the range of $a_{i} < X$ ($\Gamma_{\rm ph}$=2.187, log${\rm F_R}$=2.258, and log${\rm VI}$=1.702), 
where the misjudged rate $\eta=3/414 \simeq 0.725\%$ for FSRQs is obtained. 
Here the misjudged rate $\eta$ is a probability that an FSRQ is misclassified as a BL Lacs, which is defined as: $\eta = N_{err}/N_{F}$, 
where $N_{F}$ is the total number of FSRQs and $N_{err}$ is the number of FSRQs that are misclassified as BL Lacs at the $a_{i} <X$ range.

In order to test our hypothesis, we randomly divide FSRQs into 10 sub-samples with one sub-sample is reserved as the verification data, and the remaining 9 sub-samples are used as the training data.
Then, the proposed analysis is repeated 10 times (the 10 folds), the misjudged rate $\eta$ is repeatedly calculated 10 times.
Finally, by averaging the 10 misjudged rates, a 10-fold cross-validation\footnote{In a K-fold cross-validation, the original samples are randomly divided into K sub-samples. Among the K subsamples, one subsample is reserved as the verification data of the test model, and the remaining K-1 subsamples are used as the training data. Then, the cross-validation process is repeated K times (multiple times), and each of the K sub-samples is accurately used as the verification data. The K results resulting from the folding can then be averaged (or otherwise combined) to produce a single estimate.}
misjudged rate $\eta$ = 0.971\% is obtained for the FSRQs.
This result suggests that the zone of $a_{i} < X$ (e.g., the lowest one-sided confidence interval value for 1$\sigma$ confidence level with $\simeq$ 0.725\% false positive rate for FSRQs) can be treated as 
a ``inviable" region for the FSRQs or as a candidate zone for the BL Lacs, called ``$a_{i} < X$ candidate zone" for BL Lacs.

Finally, we can test the ``$a_{i} < X$ candidate zone'' in the BCU sample. We obtain 120 BL Lac candidates which fall into the high-confidence zone with all following three conditions satisfied: $\Gamma_{\rm ph} < 2.187$, log${\rm F_R} < 2.258$ and ${\rm logVI} < 1.702$. These 120 sources are plotted as blue solid circles in Figure \ref{sub:fig_all} (right column) and listed in Table \ref{tab:1.7zone}, while the red empty squares mark the rest unidentified optical classification BCUs.


\startlongtable
\begin{deluxetable}{lllrrrrcccccccl}
\tablecaption{The identified BL Lac candidates using the ``$a_i < X$ candidate zone'' \label{tab:1.7zone}}
\tablewidth{700pt}
\tabletypesize{\scriptsize}
\tablehead{
\colhead{3FGL name}  & \colhead{Class} & \colhead{SED}& \colhead{${\rm logF_R}$} & \colhead{$\Gamma_{\rm ph}$}& \colhead{logVI}   
&\colhead{$X$} 
& \colhead{${\rm M_8}$} & \colhead{${\rm DT_8}$} & \colhead{${\rm RF_8}$} & \colhead{${\rm SVM_8}$} & \colhead{LP17}& \colhead{Chi16}& \colhead{Y17} & \colhead{$\rm Class_{O}$}
} 
\decimalcolnumbers
\startdata	
3FGL J0003.2$-$5246	&	BCU	&	 HSP 	&	1.681 	&	1.815 	&	1.895 	&	bll	&	 bll 	&	 bll 	&	 bll 	&	 bll 	&	bll	&	bll   	&	bll	&	$-$	\\
3FGL J0017.2$-$0643	&	BCU	&	 LSP 	&	1.573 	&	1.973 	&	2.116 	&	bll	&	 bll 	&	 bll 	&	 bll 	&	 bll 	&	bll	&	bll   	&	bll	&	$-$	\\
3FGL J0031.3$+$0724	&	BCU	&	 HSP 	&	1.519 	&	1.086 	&	1.824 	&	bll	&	 bll 	&	 bll 	&	 bll 	&	 bll 	&	bll	&	bll   	&	bll	&	$bll^{e,h}$	\\
3FGL J0039.0$-$2218	&	BCU	&	 HSP 	&	1.563 	&	2.069 	&	1.715 	&	bll	&	 bll 	&	 bll 	&	 bll 	&	 bll 	&	bll	&	bll   	&	bll	&	$-$	\\
3FGL J0039.1$+$4330	&	BCU	&	 $-$ 	&	1.549 	&	0.913 	&	1.963 	&	bll	&	 bll 	&	 bll 	&	 bll 	&	 bll 	&	bll	&	bll   	&	bll	&	$-$	\\
3FGL J0040.3$+$4049	&	BCU	&	 $-$ 	&	1.481 	&	1.683 	&	1.132 	&	bll	&	 bll 	&	 bll 	&	 bll 	&	 bll 	&	bll	&	bll   	&	bll	&	$bll^{e,h}$	\\
3FGL J0040.5$-$2339	&	BCU	&	 ISP 	&	1.692 	&	1.730 	&	1.946 	&	bll	&	 bll 	&	 bll 	&	 bll 	&	 bll 	&	bll	&	bll   	&	bll	&	$bll^{g,h}$	\\
3FGL J0043.5$-$0444	&	BCU	&	 HSP 	&	1.605 	&	1.475 	&	1.735 	&	bll	&	 bll 	&	 bll 	&	 bll 	&	 bll 	&	bll	&	bll   	&	bll	&	$bll^{a,b,c,h}$	\\
3FGL J0043.7$-$1117	&	BCU	&	 HSP 	&	1.442 	&	1.397 	&	1.594 	&	bll	&	 bll 	&	 bll 	&	 bll 	&	 bll 	&	bll	&	bll   	&	bll	&	$bll^{h}$	\\
3FGL J0051.2$-$6241	&	BCU	&	 HSP 	&	1.701 	&	1.635 	&	1.663 	&	bll	&	 bll 	&	 bll 	&	 bll 	&	 bll 	&	bll	&	bll   	&	bll	&	$bll^{c,h}$	\\
3FGL J0107.0$-$1208	&	BCU	&	 ISP 	&	1.514 	&	1.778 	&	2.180 	&	bll	&	 bll 	&	 bll 	&	 bll 	&	 bll 	&	bll	&	bll   	&	bll	&	$-$	\\
3FGL J0116.2$-$2744	&	BCU	&	 $-$ 	&	1.606 	&	1.237 	&	2.023 	&	bll	&	 bll 	&	 bll 	&	 bll 	&	 bll 	&	bll	&	bll   	&	bll	&	$bll^{h}$	\\
3FGL J0121.7$+$5154	&	BCU	&	 $-$ 	&	1.586 	&	0.928 	&	1.984 	&	bll	&	 bll 	&	 bll 	&	 bll 	&	 bll 	&	bll	&	bll   	&	bll	&	$-$	\\
3FGL J0127.2$+$0325	&	BCU	&	 HSP 	&	1.695 	&	1.208 	&	1.899 	&	bll	&	 bll 	&	 bll 	&	 bll 	&	 bll 	&	bll	&	bll   	&	bll	&	$bll^{c,d,e,h}$	\\
3FGL J0139.9$+$8735	&	BCU	&	 ISP 	&	1.624 	&	1.063 	&	1.891 	&	bll	&	 bll 	&	 bll 	&	 bll 	&	 bll 	&	bll	&	bll   	&	bll	&	$-$	\\
3FGL J0150.5$-$5447	&	BCU	&	 HSP 	&	1.643 	&	1.641 	&	2.118 	&	bll	&	 bll 	&	 bll 	&	 bll 	&	 bll 	&	bll	&	bll   	&	bll	&	$-$	\\
3FGL J0156.9$-$4742	&	BCU	&	 HSP 	&	1.494 	&	1.458 	&	2.009 	&	bll	&	 bll 	&	 bll 	&	 bll 	&	 bll 	&	bll	&	bll   	&	bll	&	$bll^{a,b,c,h}$	\\
3FGL J0211.2$-$0649	&	BCU	&	 ISP 	&	1.520 	&	1.347 	&	2.100 	&	bll	&	 bll 	&	 bll 	&	 bll 	&	 bll 	&	bll	&	bll   	&	bll	&	$bll^{g}$	\\
3FGL J0213.1$-$2720	&	BCU	&	 LSP 	&	1.548 	&	1.690 	&	2.089 	&	bll	&	 bll 	&	 bll 	&	 bll 	&	 bll 	&	bll	&	bll   	&	bll	&	$-$	\\
3FGL J0228.7$-$3106	&	BCU	&	 ISP 	&	1.541 	&	1.992 	&	2.140 	&	bll	&	 bll 	&	 bll 	&	 bll 	&	 bll 	&	bll	&	bll   	&	fsrq	&	$-$	\\
3FGL J0232.9$+$2606	&	BCU	&	 ISP 	&	1.530 	&	1.993 	&	2.086 	&	bll	&	 bll 	&	 bll 	&	 bll 	&	 bll 	&	bll	&	bll   	&	bll	&	$bll^{c,h}$	\\
3FGL J0255.8$+$0532	&	BCU	&	 LSP 	&	1.602 	&	2.017 	&	2.070 	&	bll	&	 bll 	&	 bll 	&	 bll 	&	 bll 	&	bll	&	bll   	&	bll	&	$bll^{a,b,h}$	\\
3FGL J0301.8$-$2721	&	BCU	&	 LSP 	&	1.432 	&	1.722 	&	2.158 	&	bll	&	 bll 	&	 bll 	&	 bll 	&	 bll 	&	bll	&	bll   	&	bll	&	$-$	\\
3FGL J0342.6$-$3006	&	BCU	&	 LSP 	&	1.567 	&	2.195 	&	1.846 	&	bll	&	 bll 	&	 bll 	&	 bll 	&	 bll 	&	bll	&	bll   	&	bll	&	$-$	\\
3FGL J0431.6$+$7403	&	BCU	&	 HSP 	&	1.612 	&	1.485 	&	1.988 	&	bll	&	 bll 	&	 bll 	&	 bll 	&	 bll 	&	bll	&	bll   	&	bll	&	$bll^{e,h}$	\\
3FGL J0434.6$+$0921	&	BCU	&	 ISP 	&	1.691 	&	2.074 	&	2.115 	&	bll	&	 bll 	&	 bll 	&	 bll 	&	 bll 	&	bll	&	bll   	&	bll	&	$bll^{e,h}$	\\
3FGL J0439.6$-$3159	&	BCU	&	 HSP 	&	1.562 	&	1.039 	&	1.771 	&	bll	&	 bll 	&	 bll 	&	 bll 	&	 bll 	&	bll	&	bll   	&	bll	&	$-$	\\
3FGL J0506.9$-$5435	&	BCU	&	 HSP 	&	1.635 	&	1.255 	&	1.603 	&	bll	&	 bll 	&	 bll 	&	 bll 	&	 bll 	&	bll	&	bll   	&	bll	&	$bll^{c,h}$	\\
3FGL J0515.5$-$0123	&	BCU	&	 $-$ 	&	1.523 	&	1.994 	&	1.755 	&	bll	&	 fsrq 	&	 bll 	&	 bll 	&	 bll 	&	bll	&	bll   	&	bll	&	$-$	\\
3FGL J0602.8$-$4016	&	BCU	&	 HSP 	&	1.696 	&	1.873 	&	1.923 	&	bll	&	 bll 	&	 bll 	&	 bll 	&	 bll 	&	bll	&	bll   	&	bll	&	$bll^{c,h}$	\\
3FGL J0611.2$+$4323	&	BCU	&	 HSP 	&	1.633 	&	1.644 	&	2.168 	&	bll	&	 bll 	&	 bll 	&	 bll 	&	 bll 	&	bll	&	bll   	&	bll	&	$-$	\\
3FGL J0626.6$-$4259	&	BCU	&	 $-$ 	&	1.684 	&	1.149 	&	1.740 	&	bll	&	 bll 	&	 bll 	&	 bll 	&	 bll 	&	bll	&	bll   	&	bll	&	$bll^{c,d,h}$	\\
3FGL J0649.6$-$3138	&	BCU	&	 HSP 	&	1.668 	&	0.877 	&	1.729 	&	bll	&	 bll 	&	 bll 	&	 bll 	&	 bll 	&	bll	&	bll   	&	bll	&	$bll^{c,d,h}$	\\
3FGL J0652.0$-$4808	&	BCU	&	 HSP 	&	1.580 	&	1.856 	&	2.044 	&	bll	&	 bll 	&	 bll 	&	 bll 	&	 bll 	&	bll	&	bll   	&	bll	&	$-$	\\
3FGL J0730.5$-$6606	&	BCU	&	 HSP 	&	1.614 	&	1.910 	&	1.789 	&	bll	&	 bll 	&	 bll 	&	 bll 	&	 bll 	&	bll	&	bll   	&	bll	&	$bll^{a,b,c,h}$	\\
3FGL J0742.4$-$8133c	&	BCU	&	 $-$ 	&	1.404 	&	1.246 	&	1.464 	&	bll	&	 bll 	&	 bll 	&	 bll 	&	 bll 	&	$-$	&	bll   	&	bll	&	$-$	\\
3FGL J0746.9$+$8511	&	BCU	&	 HSP 	&	1.571 	&	1.057 	&	1.787 	&	bll	&	 bll 	&	 bll 	&	 bll 	&	 bll 	&	bll	&	bll   	&	bll	&	$-$	\\
3FGL J0827.2$-$0711	&	BCU	&	 HSP 	&	1.597 	&	2.241 	&	2.067 	&	bll	&	 bll 	&	 bll 	&	 bll 	&	 bll 	&	bll	&	bll   	&	bll	&	$bll^{a,b,c,d,h}$	\\
3FGL J0917.3$-$0344	&	BCU	&	 HSP 	&	1.652 	&	1.508 	&	1.764 	&	bll	&	 bll 	&	 bll 	&	 bll 	&	 bll 	&	bll	&	bll   	&	bll	&	$bll^{h}$	\\
3FGL J0921.0$-$2258	&	BCU	&	 HSP 	&	1.510 	&	1.158 	&	1.553 	&	bll	&	 bll 	&	 bll 	&	 bll 	&	 bll 	&	bll	&	bll   	&	bll	&	$bll^{a,b,c,h}$	\\
3FGL J0947.1$-$2542	&	BCU	&	 HSP 	&	1.684 	&	1.624 	&	1.950 	&	bll	&	 bll 	&	 bll 	&	 bll 	&	 bll 	&	bll	&	bll   	&	bll	&	$bll^{c,d,h}$	\\
3FGL J0953.1$-$7657c	&	BCU	&	 ISP 	&	1.567 	&	1.378 	&	1.912 	&	bll	&	 bll 	&	 bll 	&	 bll 	&	 bll 	&	$-$	&	bll   	&	bll	&	$-$	\\
3FGL J1040.8$+$1342	&	BCU	&	 HSP 	&	1.552 	&	0.748 	&	1.760 	&	bll	&	 bll 	&	 bll 	&	 bll 	&	 bll 	&	bll	&	bll   	&	bll	&	$bll^{a,b,c,h}$	\\
3FGL J1042.0$-$0557	&	BCU	&	 HSP 	&	1.618 	&	1.923 	&	1.944 	&	bll	&	 bll 	&	 bll 	&	 bll 	&	 bll 	&	bll	&	bll   	&	bll	&	$bll^{h}$	\\
3FGL J1042.1$-$4126	&	BCU	&	 HSP 	&	1.649 	&	1.310 	&	1.976 	&	bll	&	 bll 	&	 bll 	&	 bll 	&	 bll 	&	bll	&	bll   	&	bll	&	$-$	\\
3FGL J1052.8$-$3741	&	BCU	&	 ISP 	&	1.549 	&	1.803 	&	1.996 	&	bll	&	 bll 	&	 bll 	&	 bll 	&	 bll 	&	bll	&	bll   	&	bll	&	$bll^{c,g,h}$	\\
3FGL J1125.0$-$2101	&	BCU	&	 HSP 	&	1.627 	&	1.561 	&	1.784 	&	bll	&	 bll 	&	 bll 	&	 bll 	&	 bll 	&	bll	&	bll   	&	bll	&	$bll^{a,b,c,d,h}$	\\
3FGL J1141.2$+$6805	&	BCU	&	 HSP 	&	1.666 	&	1.353 	&	1.611 	&	bll	&	 bll 	&	 bll 	&	 bll 	&	 bll 	&	bll	&	bll   	&	bll	&	$-$	\\
3FGL J1141.6$-$1406	&	BCU	&	 HSP 	&	1.634 	&	1.777 	&	2.176 	&	bll	&	 bll 	&	 bll 	&	 bll 	&	 bll 	&	bll	&	bll   	&	bll	&	$bll^{b,h}$	\\
3FGL J1153.7$-$2555	&	BCU	&	 $-$ 	&	1.603 	&	1.927 	&	2.015 	&	bll	&	 bll 	&	 bll 	&	 bll 	&	 bll 	&	bll	&	bll   	&	bll	&	$-$	\\
3FGL J1155.4$-$3417	&	BCU	&	 HSP 	&	1.488 	&	1.377 	&	1.335 	&	bll	&	 bll 	&	 bll 	&	 bll 	&	 bll 	&	bll	&	bll   	&	bll	&	$bll^{f}$	\\
3FGL J1156.7$-$2250	&	BCU	&	 HSP 	&	1.584 	&	1.235 	&	1.890 	&	bll	&	 bll 	&	 bll 	&	 bll 	&	 bll 	&	bll	&	bll   	&	bll	&	$-$	\\
3FGL J1158.9$+$0818	&	BCU	&	 $-$ 	&	1.559 	&	0.869 	&	1.870 	&	bll	&	 bll 	&	 bll 	&	 bll 	&	 bll 	&	bll	&	bll   	&	bll	&	$-$	\\
3FGL J1159.6$-$0723	&	BCU	&	 LSP 	&	1.654 	&	1.887 	&	2.104 	&	bll	&	 bll 	&	 bll 	&	 bll 	&	 bll 	&	bll	&	bll   	&	bll	&	$bll^{c,h}$	\\
3FGL J1203.5$-$3925	&	BCU	&	 HSP 	&	1.602 	&	1.811 	&	1.639 	&	bll	&	 bll 	&	 bll 	&	 bll 	&	 bll 	&	bll	&	bll   	&	bll	&	$bll^{c,d,h}$	\\
3FGL J1207.6$-$4537	&	BCU	&	 $-$ 	&	1.535 	&	2.221 	&	2.113 	&	bll	&	 fsrq 	&	 bll 	&	 bll 	&	 fsrq 	&	bll	&	bll   	&	bll	&	$-$	\\
3FGL J1223.3$-$3028	&	BCU	&	 HSP 	&	1.581 	&	0.918 	&	1.887 	&	bll	&	 bll 	&	 bll 	&	 bll 	&	 bll 	&	bll	&	bll   	&	bll	&	$bll^{f}$	\\
3FGL J1258.7$+$5137	&	BCU	&	 $-$ 	&	1.611 	&	1.705 	&	2.159 	&	bll	&	 bll 	&	 bll 	&	 bll 	&	 bll 	&	bll	&	bll   	&	bll	&	$-$	\\
3FGL J1314.7$-$4237	&	BCU	&	 HSP 	&	1.653 	&	1.143 	&	2.082 	&	bll	&	 bll 	&	 bll 	&	 bll 	&	 bll 	&	bll	&	bll   	&	bll	&	$bll^{h}$	\\
3FGL J1315.4$+$1130	&	BCU	&	 HSP 	&	1.692 	&	1.316 	&	1.962 	&	bll	&	 bll 	&	 bll 	&	 bll 	&	 bll 	&	unc	&	bll   	&	bll	&	$bll^{a,b,c,h}$	\\
3FGL J1342.7$+$0945	&	BCU	&	 ISP 	&	1.373 	&	1.439 	&	1.870 	&	bll	&	 bll 	&	 bll 	&	 bll 	&	 bll 	&	bll	&	bll   	&	bll	&	fsrq 	\\
3FGL J1346.9$-$2958	&	BCU	&	 ISP 	&	1.662 	&	1.439 	&	1.744 	&	bll	&	 bll 	&	 bll 	&	 bll 	&	 bll 	&	bll	&	bll   	&	bll	&	$bll^{b,c,h}$	\\
3FGL J1356.3$-$4029	&	BCU	&	 ISP 	&	1.648 	&	1.881 	&	2.060 	&	bll	&	 bll 	&	 bll 	&	 bll 	&	 bll 	&	bll	&	bll   	&	bll	&	$-$	\\
3FGL J1406.0$-$2508	&	BCU	&	 HSP 	&	1.660 	&	1.489 	&	1.893 	&	bll	&	 bll 	&	 bll 	&	 bll 	&	 bll 	&	bll	&	bll   	&	bll	&	$bll^{b,c,h}$	\\
3FGL J1427.8$-$3215	&	BCU	&	 ISP 	&	1.557 	&	1.086 	&	2.036 	&	bll	&	 bll 	&	 bll 	&	 bll 	&	 bll 	&	bll	&	bll   	&	bll	&	$bll^{c,h}$	\\
3FGL J1434.6$+$6640	&	BCU	&	 HSP 	&	1.592 	&	0.997 	&	1.517 	&	bll	&	 bll 	&	 bll 	&	 bll 	&	 bll 	&	bll	&	bll   	&	bll	&	$bll^{b,c,h}$	\\
3FGL J1440.0$-$3955	&	BCU	&	 HSP 	&	1.610 	&	1.300 	&	1.864 	&	bll	&	 bll 	&	 bll 	&	 bll 	&	 bll 	&	bll	&	bll   	&	bll	&	$bll^{g}$	\\
3FGL J1446.8$-$1831	&	BCU	&	 HSP 	&	1.415 	&	1.409 	&	1.723 	&	bll	&	 bll 	&	 bll 	&	 bll 	&	 bll 	&	bll	&	bll   	&	bll	&	$-$	\\
3FGL J1507.6$-$3710	&	BCU	&	 ISP 	&	1.611 	&	1.898 	&	2.131 	&	bll	&	 bll 	&	 bll 	&	 bll 	&	 bll 	&	bll	&	bll   	&	bll	&	$-$	\\
3FGL J1511.8$-$0513	&	BCU	&	 $-$ 	&	1.677 	&	1.244 	&	2.034 	&	bll	&	 bll 	&	 bll 	&	 bll 	&	 bll 	&	bll	&	bll   	&	bll	&	$bll^{b,c,h}$	\\
3FGL J1512.2$-$2255	&	BCU	&	 HSP 	&	1.611 	&	1.285 	&	1.907 	&	bll	&	 bll 	&	 bll 	&	 bll 	&	 bll 	&	bll	&	bll   	&	bll	&	$bll^{c,d,h}$	\\
3FGL J1539.8$-$1128	&	BCU	&	 HSP 	&	1.612 	&	1.897 	&	2.085 	&	bll	&	 bll 	&	 bll 	&	 bll 	&	 bll 	&	bll	&	bll   	&	bll	&	$bll^{c,d,h}$	\\
3FGL J1547.1$-$2801	&	BCU	&	 HSP 	&	1.628 	&	1.677 	&	1.708 	&	bll	&	 bll 	&	 bll 	&	 bll 	&	 bll 	&	bll	&	bll   	&	bll	&	$bll^{c,d,h}$	\\
3FGL J1549.7$-$0658	&	BCU	&	 HSP 	&	1.689 	&	1.117 	&	1.924 	&	bll	&	 bll 	&	 bll 	&	 bll 	&	 bll 	&	bll	&	bll   	&	bll	&	$-$	\\
3FGL J1559.8$-$2525	&	BCU	&	 $-$ 	&	1.493 	&	1.657 	&	1.944 	&	bll	&	 bll 	&	 bll 	&	 bll 	&	 bll 	&	bll	&	bll   	&	bll	&	$bll^{g}$	\\
3FGL J1626.4$-$7640	&	BCU	&	 ISP 	&	1.692 	&	2.113 	&	1.990 	&	bll	&	 bll 	&	 bll 	&	 bll 	&	 bll 	&	bll	&	bll   	&	bll	&	$bll^{b,c,h}$	\\
3FGL J1636.7$+$2624	&	BCU	&	 ISP 	&	1.571 	&	1.312 	&	2.039 	&	bll	&	 bll 	&	 bll 	&	 bll 	&	 bll 	&	bll	&	bll   	&	bll	&	$bll^{a,b,h}$	\\
3FGL J1643.6$-$0642	&	BCU	&	 HSP 	&	1.694 	&	1.459 	&	2.071 	&	bll	&	 bll 	&	 bll 	&	 bll 	&	 bll 	&	bll	&	bll   	&	bll	&	$bll^{h}$	\\
3FGL J1656.8$-$2010	&	BCU	&	 HSP 	&	1.680 	&	1.572 	&	1.961 	&	bll	&	 bll 	&	 bll 	&	 bll 	&	 bll 	&	bll	&	bll   	&	bll	&	$bll^{a,b,c,d,h}$	\\
3FGL J1711.6$+$8846	&	BCU	&	 $-$ 	&	1.587 	&	1.077 	&	1.570 	&	bll	&	 bll 	&	 bll 	&	 bll 	&	 bll 	&	bll	&	bll   	&	bll	&	$-$	\\
3FGL J1714.1$-$2029	&	BCU	&	 HSP 	&	1.655 	&	0.931 	&	1.344 	&	bll	&	 bll 	&	 bll 	&	 bll 	&	 bll 	&	bll	&	bll   	&	bll	&	$-$	\\
3FGL J1716.7$-$8112	&	BCU	&	 HSP 	&	1.623 	&	2.028 	&	2.060 	&	bll	&	 bll 	&	 bll 	&	 bll 	&	 bll 	&	bll	&	bll   	&	bll	&	$bll^{h}$	\\
3FGL J1719.3$+$1206	&	BCU	&	 ISP 	&	1.483 	&	1.760 	&	2.078 	&	bll	&	 bll 	&	 bll 	&	 bll 	&	 bll 	&	bll	&	bll   	&	bll	&	$-$	\\
3FGL J1735.4$-$1118	&	BCU	&	 LSP 	&	1.659 	&	1.926 	&	2.156 	&	bll	&	 bll 	&	 bll 	&	 bll 	&	 bll 	&	bll	&	fsrq     	&	bll	&	$-$	\\
3FGL J1740.4$+$5347	&	BCU	&	 LSP 	&	1.616 	&	1.693 	&	2.019 	&	bll	&	 bll 	&	 bll 	&	 bll 	&	 bll 	&	bll	&	bll   	&	bll	&	$bll^{c,h}$	\\
3FGL J1757.1$+$1533	&	BCU	&	 LSP 	&	1.653 	&	2.256 	&	2.045 	&	bll	&	 bll 	&	 bll 	&	 bll 	&	 fsrq 	&	bll	&	bll   	&	bll	&	$-$	\\
3FGL J1820.3$+$3625	&	BCU	&	 HSP 	&	1.546 	&	1.143 	&	1.777 	&	bll	&	 bll 	&	 bll 	&	 bll 	&	 bll 	&	bll	&	bll   	&	bll	&	$bll^{e,h}$	\\
3FGL J1824.4$+$4310	&	BCU	&	 ISP 	&	1.443 	&	1.528 	&	1.725 	&	bll	&	 bll 	&	 bll 	&	 bll 	&	 bll 	&	bll	&	bll   	&	bll	&	$-$	\\
3FGL J1838.5$-$6006	&	BCU	&	 HSP 	&	1.639 	&	1.901 	&	1.857 	&	bll	&	 bll 	&	 bll 	&	 bll 	&	 bll 	&	bll	&	bll   	&	bll	&	$-$	\\
3FGL J1841.2$+$2910	&	BCU	&	 HSP 	&	1.604 	&	1.811 	&	1.567 	&	bll	&	 bll 	&	 bll 	&	 bll 	&	 bll 	&	bll	&	bll   	&	bll	&	$bll^{e,h}$	\\
3FGL J1848.1$-$4230	&	BCU	&	 $-$ 	&	1.598 	&	2.009 	&	1.951 	&	bll	&	 bll 	&	 bll 	&	 bll 	&	 bll 	&	bll	&	bll   	&	bll	&	$-$	\\
3FGL J1855.1$-$6008	&	BCU	&	 $-$ 	&	1.499 	&	1.922 	&	1.813 	&	bll	&	 bll 	&	 bll 	&	 bll 	&	 bll 	&	bll	&	bll   	&	bll	&	$bll^{c,h}$	\\
3FGL J1904.5$+$3627	&	BCU	&	 HSP 	&	1.583 	&	1.995 	&	2.098 	&	bll	&	 bll 	&	 bll 	&	 bll 	&	 bll 	&	bll	&	bll   	&	bll	&	$bll^{e,h}$	\\
3FGL J1913.9$+$4441	&	BCU	&	 HSP 	&	1.481 	&	1.192 	&	1.851 	&	bll	&	 bll 	&	 bll 	&	 bll 	&	 bll 	&	bll	&	bll   	&	bll	&	$bll^{b,c,h}$	\\
3FGL J1939.6$-$4925	&	BCU	&	 $-$ 	&	1.621 	&	1.009 	&	1.624 	&	bll	&	 bll 	&	 bll 	&	 bll 	&	 bll 	&	bll	&	bll   	&	bll	&	$-$	\\
3FGL J1944.1$-$4523	&	BCU	&	 HSP 	&	1.591 	&	1.828 	&	1.560 	&	bll	&	 bll 	&	 bll 	&	 bll 	&	 bll 	&	bll	&	bll   	&	bll	&	$-$	\\
3FGL J1954.9$-$5640	&	BCU	&	 HSP 	&	1.644 	&	0.924 	&	1.878 	&	bll	&	 bll 	&	 bll 	&	 bll 	&	 bll 	&	bll	&	bll   	&	bll	&	$bll^{g}$	\\
3FGL J1955.0$-$1605	&	BCU	&	 HSP 	&	1.494 	&	1.394 	&	2.047 	&	bll	&	 bll 	&	 bll 	&	 bll 	&	 bll 	&	bll	&	bll   	&	bll	&	$bll^{a,b,c,d,,h}$	\\
3FGL J1955.9$+$0212	&	BCU	&	 $-$ 	&	1.607 	&	1.507 	&	1.927 	&	bll	&	 bll 	&	 bll 	&	 bll 	&	 bll 	&	bll	&	bll   	&	bll	&	$bll^{c,h}$	\\
3FGL J1959.8$-$4725	&	BCU	&	 HSP 	&	1.695 	&	1.377 	&	1.524 	&	bll	&	 bll 	&	 bll 	&	 bll 	&	 bll 	&	bll	&	bll   	&	bll	&	$bll^{c,h}$	\\
3FGL J2002.7$+$6303	&	BCU	&	 LSP 	&	1.634 	&	1.065 	&	2.127 	&	bll	&	 bll 	&	 bll 	&	 bll 	&	 bll 	&	bll	&	bll   	&	bll	&	$-$	\\
3FGL J2014.5$+$0648	&	BCU	&	 HSP 	&	1.650 	&	1.203 	&	1.915 	&	bll	&	 bll 	&	 bll 	&	 bll 	&	 bll 	&	bll	&	bll   	&	bll	&	$bll^{c,h}$	\\
3FGL J2017.6$-$4110	&	BCU	&	 HSP 	&	1.590 	&	2.159 	&	2.160 	&	bll	&	 bll 	&	 bll 	&	 bll 	&	 bll 	&	bll	&	bll   	&	F	&	$-$	\\
3FGL J2026.3$+$7644	&	BCU	&	 HSP 	&	1.505 	&	0.805 	&	1.839 	&	bll	&	 bll 	&	 bll 	&	 bll 	&	 bll 	&	bll	&	bll   	&	bll	&	$-$	\\
3FGL J2031.0$+$1937	&	BCU	&	 HSP 	&	1.671 	&	1.754 	&	1.826 	&	bll	&	 bll 	&	 bll 	&	 bll 	&	 bll 	&	bll	&	bll   	&	bll	&	$bll^{b,c,h}$	\\
3FGL J2036.6$-$3325	&	BCU	&	 HSP 	&	1.443 	&	0.547 	&	1.305 	&	bll	&	 bll 	&	 bll 	&	 bll 	&	 bll 	&	unc	&	bll   	&	bll	&	$bll^{b,c,h}$	\\
3FGL J2046.7$-$1011	&	BCU	&	 HSP 	&	1.550 	&	2.026 	&	1.609 	&	bll	&	 bll 	&	 bll 	&	 bll 	&	 bll 	&	bll	&	bll   	&	bll	&	$bll^{g}$	\\
3FGL J2104.2$-$0211	&	BCU	&	 HSP 	&	1.528 	&	1.150 	&	1.524 	&	bll	&	 bll 	&	 bll 	&	 bll 	&	 bll 	&	bll	&	bll   	&	bll	&	$bll^{a,b,c,h}$	\\
3FGL J2133.3$+$2533	&	BCU	&	 ISP 	&	1.608 	&	1.600 	&	2.010 	&	bll	&	 bll 	&	 bll 	&	 bll 	&	 bll 	&	bll	&	bll   	&	bll	&	$bll^{h}$	\\
3FGL J2212.6$+$2801	&	BCU	&	 LSP 	&	1.600 	&	2.161 	&	1.791 	&	bll	&	 bll 	&	 bll 	&	 bll 	&	 bll 	&	bll	&	bll   	&	bll	&	$bll^{e,g,h}$	\\
3FGL J2213.6$-$4755	&	BCU	&	 $-$ 	&	1.562 	&	1.436 	&	1.889 	&	bll	&	 bll 	&	 bll 	&	 bll 	&	 bll 	&	bll	&	bll   	&	bll	&	$bll^{g}$	\\
3FGL J2220.3$+$2812	&	BCU	&	 HSP 	&	1.579 	&	1.689 	&	1.833 	&	bll	&	 bll 	&	 bll 	&	 bll 	&	 bll 	&	bll	&	bll   	&	bll	&	$bll^{e,h}$	\\
3FGL J2232.9$-$2021	&	BCU	&	 HSP 	&	1.511 	&	1.099 	&	2.081 	&	bll	&	 bll 	&	 bll 	&	 bll 	&	 bll 	&	bll	&	bll   	&	bll	&	$bll^{a,b,h}$	\\
3FGL J2243.2$-$3933	&	BCU	&	 LSP 	&	1.530 	&	1.773 	&	2.119 	&	bll	&	 bll 	&	 bll 	&	 bll 	&	 bll 	&	bll	&	bll   	&	bll	&	$-$	\\
3FGL J2251.5$-$4928	&	BCU	&	 ISP 	&	1.689 	&	1.522 	&	1.967 	&	bll	&	 bll 	&	 bll 	&	 bll 	&	 bll 	&	bll	&	bll   	&	bll	&	$bll^{c,h}$	\\
3FGL J2305.3$-$4219	&	BCU	&	 LSP 	&	1.614 	&	1.600 	&	2.048 	&	bll	&	 bll 	&	 bll 	&	 bll 	&	 bll 	&	bll	&	bll   	&	bll	&	$bll^{g}$	\\
3FGL J2312.9$-$6923	&	BCU	&	 $-$ 	&	1.581 	&	1.086 	&	1.804 	&	bll	&	 bll 	&	 bll 	&	 bll 	&	 bll 	&	bll	&	bll   	&	bll	&	$-$	\\
3FGL J2316.8$-$5209	&	BCU	&	 ISP 	&	1.543 	&	1.408 	&	1.735 	&	bll	&	 bll 	&	 bll 	&	 bll 	&	 bll 	&	bll	&	bll   	&	bll	&	$bll^{g}$	\\
3FGL J2322.9$-$4917	&	BCU	&	 HSP 	&	1.634 	&	1.452 	&	1.957 	&	bll	&	 bll 	&	 bll 	&	 bll 	&	 bll 	&	bll	&	bll   	&	bll	&	$bll^{c,h}$	\\
3FGL J2353.3$-$4805	&	BCU	&	 $-$ 	&	1.457 	&	1.947 	&	2.011 	&	bll	&	 fsrq 	&	 bll 	&	 bll 	&	 bll 	&	bll	&	bll   	&	bll	&	$-$	\\
\enddata
\end{deluxetable}
\tablecomments{Column 1 shows the 3FGL names.
Column 2  lists the optical class (BCU reported in \citealt{2015ApJS..218...23A}).
Column 3 gives the SED classifications (LSP, ISP and HSP);
the radio flux  (log${\rm F_R}$) is listed in Column 4.
The $\gamma$-ray photon spectral index ($\Gamma_{\rm ph}$) and $\gamma$-ray variability index (log${\rm VI}$) and  are shown in Columns 5 and 6, respectively.
The BL Lac candidates using the ``$a_i < X$ candidate zone''  are listed in  Column 7.
Columns 8-11 ({$\rm M_8$}, {$\rm DT_8$}, {$\rm RF_8$}, and $\rm SVM_8$) indicate the BL Lac - type (``bll") candidates identified by 4 different supervised machine learning (SML) algorithms (Mclust Gaussian finite mixture models ($\rm M_8$), Decision trees ($\rm DT_8$), Random forests ($\rm RF_8$) and support vector machines ($\rm SVM_8$)) with 8 parameters  in \citealt{2019ApJ...872..189K}.
Column 12 (LP17) lists the classifications (``bll" for BL Lac,  ``{unc}" for uncertain and ``-" for a mismatched source by cross comparison) in  \citealt{2017A&A...602A..86L} using multivariate classifications.
Column 13 (Chi16) reports the classifications in  \citealt{2016MNRAS.462.3180C} using artificial neural networks (ANN) machine-learning techniques.
Column 14 (Y17) shows the identified BL Lacs reported in \citealt{2017ApJ...838...34Y} by researching the spectral index. 
Column 15  ($\rm Class_{O}$) reports the optical classification (identified BL Lac types based on optical spectroscopic)
in \citealt{2016Ap&SS.361..316A} ($^a$),
\citealt{2016Ap&SS.361..337M} ($^b$); 
\citealt{2017ApJS..232...18A} ($^c$);
\citealt{2017Ap&SS.362..228P} ($^d$) 
\citealt{2018AJ....156..212M} ($^e$);
\citealt{2019ApJS..241....5D} ($^f$); 
\citealt{2019Ap&SS.364....5M} ($^g$);
or/and
\citealt{2019arXiv190210045T}($^h$),
 respectively.}

\section{Comparison with literature  results}\label{sec_comp}

We then compared our 120 identified BL Lac candidates with some other recent studies. We found that our results are mostly consistent with previous works presented in {\cite{2016MNRAS.462.3180C}; \cite{2017A&A...602A..86L}; \cite{2017ApJ...838...34Y} and \cite{2019ApJ...872..189K}} which utilize different statistical (e.g., SML) algorithms (see Table \ref{tab_ccomparison} and Table \ref{tab:1.7zone}). The exceptions are as follows: 2 sources do not find matching sources and 2 sources did not provide a clear classification in \cite{2017A&A...602A..86L}. In addition, only 3 sources are classified as FSRQs in \emph{$Mclust$ Gaussian Mixture Modelling} (${\rm M_8}$), and two are classified as FRSQs using \emph{support vector machine} (${\rm SVM_8}$) using 8 parameters in \citealt{2019ApJ...872..189K}; 
1 source is classified as an FSRQ in \citealt{2016MNRAS.462.3180C} (Chi16), whereas two sources are classified as FRSQs in \citealt{2017ApJ...838...34Y} (Y17).
The results, provided in Table \ref{tab_ccomparison}, indicate the highest mismatch rate (e.g., rate = 3/120\% $\sim$2.5\%) is less than 3\%. Hence, the selected area ($a_i< X$ candidate zone) shows a higher degree of confidence.

For these 120 identified BL Lac candidates in the work, 
of which 41 sources are identified as BL Lac-type in the 3FHL catalog (\citealt{2017ApJS..232...18A}, 3FHL, see Table \ref{tab_ccomparison});
and 63 sources are identified as BL Lac-type in 4FGL catalog
(see 4FGL FITS table ``gll\_psc\_v20.fit"\footnote{\url{https://fermi.gsfc.nasa.gov/ssc/data/access/lat/8yr_catalog/gll_psc_v20.fit}} of \citealt{2019arXiv190210045T}, 4FGL).
Only 1 source is classified as an FSRQ in the 4FGL catalog \citep{2019arXiv190210045T}.
There are 24, 2, 11, 10, 12, and 15 sources
that have been identified as the BL Lac-type by 
\cite{2016Ap&SS.361..337M} (M16),
\cite{2019ApJS..241....5D} (D19), 
\cite{2019Ap&SS.364....5M} (M19),
\cite{2018AJ....156..212M} (M18),
\cite{2017Ap&SS.362..228P} (P17), and
\cite{2016Ap&SS.361..316A} (A16) using spectroscopic observations, respectively. 
After cross-matching these results (3FHL, M16, D19, M19, M18, P17, A16, and 4FGL), 74 sources are obtained (also see Table \ref{tab_ccomparison} and Table \ref{tab:1.7zone}).
Here, the remaining ones (``46 sources") need to be further tested and confirmed by spectroscopic observations.

\begin{table*}[tbp]
	\centering
\caption{The comparison results}\label{tab_ccomparison}
\begin{tabular}{c|cccccccc|cccccccc}
 \hline \hline
 Class & $N_{X}$ &${\rm M_8}$ & ${\rm DT_8}$ & ${\rm RS_8}$ &	${\rm SVM}_{8}$ & Chi16 &{LP17} &	 Y17	&	{M16} & 3FHL & D19 & M19 & M18 & P17 & A16 &4FGL \\
 \colnumbers \\
 $-$ & & & 	& 	&	 	&	 	 &	2 	&	 	&	 96 & 36 & 118 & 109 & 110 & 108 & 105 & 4 \\ 
bll & 120 &117 &	120	&	120	 &	118 	&	119 	&	116 	&	118	&	{24} & 41 & 2 & 11 & 10 & 12 & 15 & 63 \\
\cline{10-17}
 & & &		&	 	&	 	&	 	&	 	& 	&\multicolumn{8}{c}{74} \\
fsrq & 0 &3 &	0	&	0 &	2 	&	1 	&	0 	&	2	&	 0 & 0 & 0 & 0 & 0 & 0 & 0 & 1 \\
unc & & &		&	 	&	 	&	 	&	2 	&		&	 & 43 & & & & & & 52 \\
\hline 
\end{tabular}
\tablecomments{
Column 1 shows the classifications ($-$ represents the number of mismatch by cross comparison, ``{bll}" ,``{fsrq}" and ``{unc}" indicate BL Lac, FSRQ and uncertain type respectively).
Column 2 is the number of sources ($N_{X}$) obtained by $a_i< X$ candidate zone.
The comparison results of \emph{$Mclust$ Gaussian Mixture Modelling (${\rm M_8}$), decision tree (${\rm DT_8}$), random forest (${\rm RS_8}$), and support vector machine (${\rm SVM_8}$)} using 8 parameters 
in \citealt{2019ApJ...872..189K} are listed in Column 3 - 6, respectively.
The results of cross comparison with \citealt{2016MNRAS.462.3180C} (Chi16); \citealt{2017A&A...602A..86L} (LP) and
 \citealt{2017ApJ...838...34Y} (Y17) are shown in Column 7, 8 and 9, respectively.
 Columns 10-17 exhibit the results of cross comparison with 
\citealt{2016Ap&SS.361..337M} (M16); 
\citealt{2017ApJS..232...18A} (3FHL);
\citealt{2019ApJS..241....5D} (D19); 
\citealt{2019Ap&SS.364....5M} (M19); 
\citealt{2018AJ....156..212M} (M18);
\citealt{2017Ap&SS.362..228P} (P17);
\citealt{2016Ap&SS.361..316A} (A16) and
\citealt{2019arXiv190210045T} (4FGL), respectively,
 where, 74 sources (also see Columns 18 in Table \ref{tab:1.7zone}) are obtained from cross-matching these results (M16, 3FHL, D19, M19, M18, P17, A16, and 4FGL).
 }
\end{table*}

\section{Discussions}\label{sec:disc}

As shown in Paper I or other similar works, the SML method can return the probabilities $P_{Bi}$ and $P_{Fi}$ (e.g., see the machine-readable supplementary material in Table 4 in \citealt{2019ApJ...872..189K}) that a BCU $i$ belongs to the BL Lacs (B) or FSRQs (F) classes, respectively. 
These probabilities can help to distinguish each source belonging to each class. 
However, it should be noted that SML algorithms provides a statistical approach (or other statistical algorithms) to address the potential classification of BCUs, but the test error rate $>$ 0.11 (e.g., Paper I) is still very large. Due to the very large misclassified value, FSRQs and BL Lacs may be misclassified. 
A more efficient (high confidence) method for evaluating the potential classification of the BCUs may be necessary, and needs to be further addressed. 
On the other hand, in fact, in this work, our aim is to obtain a more precise conclusion with the least, most direct observation with the simplest method. 
Although there is still some artificiality in limiting the boundary value of ``$a_i< X$ candidate zone", the result of ``$ a_i <X $candidate zone'' is stable.
Here, only a part of BL Lacs are classified from the BCUs, but not the majority. 
The results likely provide some clues to the further study.
For instance, it can contribute to subsequent source selections in the spectroscopic observation campaigns needed to confirm their real nature and, possibly, 
determine their redshifts
(see, e.g., \citealt{2014ApJ...780...73A}), perform population studies of the remaining unassociated $\gamma$-ray sources (e.g., see \citealt{2013ApJ...779..133A}; 
\citealt{2019ApJS..242....4D} for some discussions).
The result of this work may provide more samples for studying the jet physics of on the population of HSP BL Lacs, 
or some clues for the planning of the main targets for rigorous analyses and multi-wavelength observational campaigns (e.g., \citealt{2019ApJ...887..104C}).
The empirical candidates zone gives higher confidence results with higher probabilities for $P_{Bi}$ (see Table 4 in paper I) that a BCU $i$ belongs to BL Lacs (B) classes.
This can provide the observer with guidance on the selection of the observation target within the limited observation resources (e.g., observation equipment, time).
However, the empirical method may still cause misjudgments in identifying the potential (optical) classification of blazars.
The optical spectroscopic observations remains the most efficient and accurate way to determine the real nature of these sources.

For the 120 predicted BL Lac candidates using the ``$a_i< X$ candidate zone" in the work, 
we also test the independence between the known classification 414 FSRQs using the two sample test.
The distributions of $\Gamma_{\rm ph}$, ${\rm VI}$ and ${\rm F_R}$ between the 414 FSRQs and the 120 identified BL Lac candidates groups are significantly different. 
The two-sample Kolmogorov$-$Smirnov test gives D = 0.725, and the p-value $ p_1 = 0$ ;
the Welch Two Sample t-test gives t = 38, df = 926, and the p-value $ p_2 <$ 1.0E-6;
while the Wilcoxon rank sum test with a continuity correction gives W = 416470 and the p-value $ p_3 <$ 1.0E-6 for all 3 of the parameters (see Table \ref{tab:test}). 
For other (one or two) parameter combinations, the test results are also reported in Table \ref{tab:test}.
Which indicates that there is a strong separation between the 120 predicted BL Lac candidates and the known classification 414 FSRQ, which further verifies our results from another perspective.

\begin{table*}[tbp]
	\centering
	\caption{Source numbers in different boundary conditions}
	\label{tab:result}
	\begin{tabular}{lcc} 
		\hline
                    {Paramaters}        & $\eta_{1\sigma}$         & $N_{1\sigma}$    \\
 \colnumbers\\
		\hline
$\Gamma_{\rm ph}$, ${\rm log VI}$, log${\rm F_R}$ 	 &(3/414)          &120 \\ 
$\Gamma_{\rm ph}$, ${\rm log F_R}$                            &(10/414)         &177 \\
$\Gamma_{\rm ph}$, ${\rm log VI}$                               &(14/414)        &142 \\
${ \rm log VI}$, log${\rm F_R}$                                      &(10/414) 	 &175 \\
$\Gamma_{\rm ph}$                                                     &(66/414)	 &212 \\
log${\rm F_R}$                                                            &(65/414) 	 &289 \\
${\rm log VI}$                                                              &(48/414) 	 &222 \\
 		\hline
	\end{tabular}\\
\tablecomments{Column 1 shows the parameters satisfied simultaneously.
Column 2 is the misjudged rate ($\eta_{1\sigma}$) in the boundary value with a one-sided confidence interval for the 1$\sigma$ confidence level.
Column 3 is the number of BL Lac candidates ($N_{1\sigma}$) selected from the BCUs in the boundary value with one-sided confidence interval for the 1$\sigma$ confidence level.}
\end{table*}

We should note that, in Figure \ref{sub:fig_all} (right column), if only two premises should be satisfied simultaneously, it would be that more sources can be selected as possible BL Lac candidates. 
For example, considering the lower, middle, and upper panels of Figure \ref{sub:fig_all}, 
there are an extra 57 BCUs with a misjudged rate (a probability that FSRQs are misclassified as BL Lacs) $\eta = 10/414 \simeq 2.415\%$ (see Table \ref{tab:result}) in the range ($\Gamma_{\rm ph}<2.187$ and ${\rm logF_R <2.258}$) in the $\Gamma_{\rm ph}-{\rm log F_R}$ panel (the lower panel of right column in Figure \ref{sub:fig_all}).
There are an extra 22 BCUs with a misjudged rate $\eta = 14/414 \simeq 3.382\%$ (see Table \ref{tab:result}) in the range ($\Gamma_{\rm ph}<2.187$ and ${ \rm log VI <1.702}$) in the $\Gamma_{\rm ph}-{\rm log VI}$ panel (the middle panel of right column in Figure \ref{sub:fig_all}).
There are an extra 55 BCUs (obtained easily from the the 3LAC Website version) with a misjudged rate $\eta = 10/414 \simeq 2.415\%$ (see Table \ref{tab:result}) in the range (${\rm log VI <1.702}$ and ${\rm log F_R <2.258}$) in the ${\rm log F_R}-{ \rm log VI}$ panel (the upper panel of right column in Figure \ref{sub:fig_all}). 
These sources (57, 22 and 55) have a larger misjudged rate ($\eta > 2.4\%$); although we did not conclusively evaluate their potential classifications (FSRQs and BL Lacs), it may be helpful for source selection in the spectroscopic observation campaigns in the future to further diagnose their optical classifications (see e.g., \citealt{2017ApJ...838...34Y,2013ApJS..207...16M} for the some discussions). In addition, if only one parameter is considered, a bigger misjudged error is introduced (see Table \ref{tab:result}).
Whether these 3 parameters ($\Gamma_{\rm ph}$, logVI and ${\rm log F_R}$) are the optimum combination of parameters needs to be further tested.

In addition, it must be highlighted that, in this work, the selection effects should be cautious (e. g., sample and method. see \citealt{2018RAA....18...56K,2019ApJ...872..189K} for the detail discussions), which may affect the source distributions and the results of the analysis.
However, this work provides a simple direct method to distinguish the BL Lacs from the BCUs based on the direct observational data.
As the expansion of the sample, whether the proposed analysis ($a_i < X$ candidate zone) in this work is always robust and effective, that uses a large and complete sample (e.g., the upcoming 4LAC) is needed to further test and address the issue.

\section{summary}\label{sec:summary}

In this work, we proposed an analysis to evaluate the potential optical classification of BCUs. 
Based on the 3LAC Clean Sample, we collect 1418 Fermi blazars with 3 parameters of photon spectral index, variability index, and radio flux. 
We study the distributions of the FSRQs and BL Lacs based on the scatterplots of these 3 parameters. 
We find that there are almost no FSRQs falling in a range: $\Gamma_{\rm ph} < a_{1}$, log${\rm F_R} < a_{2}$ and ${ \rm logVI} < a_{3}$ for these 3 parameters.
However some BL Lacs lie in the zone ($a_{i} < X$). Therefore, we suggest that it may be an invalid zone for FSRQs, but may be a candidate zone for BL Lacs (called ``$a_i < X$ candidate zone'' for BL Lacs). 
Using one-sample normal distribution tests for the $\Gamma_{\rm ph}$, ${\rm VI}$, and ${\rm F_R}$ of the FSRQs, which show that these three variables have normal distributions. We assume that the lowest one-sided confidence interval values are treated as the boundary values $a_{i}$ of these three parameters. 
In the unilateral 1$\sigma$ confidence level, $a_{1}=2.187$, $a_{2}=2.258$ and $a_{3}=1.702$ are calculated.
Assuming $\Gamma_{\rm ph} < 2.187$, log${\rm F_R} < 2.258$ and ${\rm logVI} < 1.702$ are satisfied simultaneously, 
we apply the ``$a_{i} < X$ candidate zone'' to the BCUs, and then obtain 120 potential BL Lac candidates. 
We compared the 120 potential BL Lac candidates with some other recent (statistical) results, 
and find that almost all of the results are consistent with the results that have been identified as BL Lacs in SML (or other statistical) methods.
We also compared the 120 potential BL Lac candidates with other spectroscopic certification results,
and find most of the 120 (74) sources have been identified as BL Lacs by spectroscopic observations (see Table 3 and Table 4).
Therefore, we suggest that the empirical candidates zone ($a_i < X$) may be a good criterion (high-confidence) for evaluating BL Lacs candidates only based on the direct observational data of $\Gamma_{\rm ph}$, VI and ${\rm F_R}$.
Although the proposed approach only identifies a part of BL Lac candidates in the BCUs, not the majority. The results are stable and with a higher degree of confidence.

\section*{Acknowledgements}
We thank the anonymous referee for very constructive and helpful comments and suggestions, which greatly helped us to improve our paper.
This work is partially supported by the National Natural Science Foundation of China (Grant Nos.11763005, 11873043, 11847091, and U1931203),
the Science and Technology Foundation of Guizhou Province (QKHJC[2019]1290),
the Research Foundation for Scientific Elitists of the Department of Education of Guizhou Province (QJHKYZ[2018]068),
the Open Fund of Guizhou Provincial Key Laboratory of Radio Astronomy and Data Processing (KF201811),
the Science and Technology Platform and Talent Team Project of Science and Technology Department of Guizhou Province (QKH $\ast$ Platform \& Talent[2018]5777, [2017]5721),
and the Research Foundation of Liupanshui Normal University (LPSSYKJTD201901,LPSSY201401, LPSSYSSDPY201704, LPSZDZY2018-03, LPSSYZDXK201801, and LPSSYsyjxsfzx201801).


 \end{CJK*}
\end{document}